\documentclass{sig-alternate-05-2015}
\usepackage{algorithm2e}
\usepackage{float}
\usepackage{enumitem}
\usepackage{tikz}
\usepackage{url}
\usepackage{amsmath,amsfonts,amssymb}
\usetikzlibrary{arrows,positioning,shapes.geometric, automata}
\begin{document}

\setcopyright{acmcopyright}

%

\title{A recommender system for efficient discovery of new anomalies in large-scale access logs}
\numberofauthors{3} 
\author{
\alignauthor Heju Jiang\textsuperscript{*} \\
       \affaddr{Instart Logic, Inc.}\\
       \email{hjiang@instartlogic.com}
\alignauthor
Scott Algatt \\
       \affaddr{Instart Logic, Inc.}\\
       \email{salgatt@instartlogic.com}
\alignauthor Parvez Ahammad\titlenote{Corresponding authors}\\
       \affaddr{Instart Logic, Inc.}\\
       \email{parvez@ieee.org}
}

\maketitle
\begin{abstract}
We present a novel, non-standard recommender system for large-scale security policy management(SPM). Our system \emph{Helios} discovers and recommends unknown and unseen anomalies in large-scale access logs with minimal supervision and no starting information on users and items. Typical recommender systems assume availability of user- and item-related information, but such information is not usually available in access logs. To resolve this problem, we first use discrete categorical labels to construct categorical combinations from access logs in a bootstrapping manner. Then, we utilize rank statistics of entity rank and order categorical combinations for recommendation. From a double-sided cold start, with minimal supervision, \emph{Helios} learns to recommend most salient anomalies at large-scale, and provides visualizations to security experts to explain rationale behind the recommendations. Our experiments show \emph{Helios} to be suitable for large-scale applications: from cold starts, in less than 60 minutes, \emph{Helios} can analyze roughly 4.6 billion records in logs of 400GB with about 300 million potential categorical combinations, then generate ranked categorical combinations as recommended discoveries. We also show that, even with limited computing resources, \emph{Helios} accelerates unknown and unseen anomaly discovery process for SPM by 1 to 3 orders of magnitude, depending on use cases. In addition, \emph{Helios'} design is flexible with metrics and measurement fields used for discoveries and recommendations. Overall, our system leads to more efficient and customizable SPM processes with faster discoveries of unseen and unknown anomalies.
\end{abstract}

%
%
%
%
%
%

\keywords{Security, Security Policy Management, Cold Start, Recommender System, Large-scale, 
Discovery, Categorical Combinations, Ranking, Mean Reciprocal Rank}

\section{INTRODUCTION \& Motivation}
Large-scale security policy management(SPM) typically involves human security experts working to refine security policies, and a major challenge is: unseen potential threats may have already evaded intrusion detection systems(IDS) and bypassed web application firewalls(WAF) \cite{waf:xss:difficult, waf:sqlinjection, waf:sqlinjection:bypasswaf, xu2016automatically}, and are recorded in web access logs, and such new and unknown anomalous patterns need to be discovered. However, without a fast discovery engine, human security experts are trapped by trying to find needles in ever growing stacks of web traffic information. The end-to-end time-to-discover(E2E TTD) is long, the process manual and exhausting, the information load vast. For example, WAF SPM is still a largely manual and slow process of analyzing web access logs to identify new anomalous patterns to find potential threats. Further, as web traffic grows, it is more critical to discover lurking anomalies from heavy loads of logs for security purposes. Hence, we realize there is a strong need to discover and prioritize new and previously unseen and unknown anomalies for security experts\cite{landwehr2012privacy, Landwehr:2008keynote}. 

A natural candidate for discovery and prioritization, is a recommender system. Recommender systems find and recommend items of interest to users. However, most current recommender systems have two major problems. First, they normally require some combinations of the following information\cite{Sarwar:item1, Deshpande:item2, Hu:implicit, Shi:ltrcf, WeiKarLeSmo08:cofirank, Shi:lessismore}: (1)User metadata(e.g. preferred categories of items); (2)Implicit data(e.g. users' binary responses to items such as ``liked'' vs. ``not liked'' instead of numerical rating scores), and (3)Pre-determined similarity measurement(e.g. how similar or different two items are to each other), but none of such information is available in common web access log formats(Sec.3.1). Second, despite the needed information being absent in the data source, mechanisms of many recommender systems rely on such information to fulfill the purpose of recommendation - for example, inferring and recommending items to users, based on users' metadata and implicit feedback to similar items. 

However, when discovering new unknown and unseen anomalies, security experts are the users, and more often than not, they do not possess a prior knowledge of unseen anomalies which are about to be discovered. It is even more unlikely that security experts could give any feedback on how much they find certain unseen, unknown, and undiscovered anomalies suitable for security policy sets. Further, human experts' preferences for new anomalies - if any at all - are not recorded in access log formats, leaving common recommender systems unable to derive preferences and similarity measurements. When anomalous patterns are discovered from post-filter access logs, human security experts would then decided if certain security policies should be amended, but at the start of the discovery process, they would not even now what to expect. 

With \emph{Helios}, we present a novel and interpretable recommender system, which achieves the following: \\
\begin{enumerate}[nosep]
\item It does not assume or rely on the availability of user metadata, implicit data, and pre-determined similarity metrics to start the process of recommending items to human experts. Detailed system design in Sec.3, Fig.1, and Fig.2. 
\item It reduces E2E TTD of unknown anomalies: our experiment consistently shows \emph{Helios} reduces E2E TTD by human experts by 1 to 3 orders of magnitude. Experiments in Sec.4. 
\item Under minimal supervision(that is, minimal amount of labeled data), \emph{Helios} can deal with cold starts, making it suitable for exploring unknown and unseen anomalies. Details in Sec.3 and 4. 
\item \emph{Helios} is agnostic to data sources, and can be broadly generalized: users can choose arbitrary categories, different categorical labels, different classes of entities to be measured and ranked, measurement metrics, and ranking functions to build customized anomaly discovery process. 
\end{enumerate} 
\vspace{0.1in}
As a minimally supervised recommender system for anomaly discovery and recommendation, \emph{Helios} has 3 distinct stages of computation(Fig.1). First, \emph{Helios} constructs categorical combinations from discrete categorical labels in access logs, then use them to compute rank statistics. Second, based on rank statistics, \emph{Helios} recommends categorical combinations where highly abnormal patterns occur. Third, \emph{Helios} supplies visualizations to human security experts for interpreting rationales behind the recommendations. Security experts can then decided if and which categorical combinations should be incorporated into existing security policy sets(e.g. access rules configured on WAF). The steps are elaborated in following sections. 
\begin{figure} 
\centering
\begin{tikzpicture}
 \tikzset{block/.style= {draw, rectangle, align=left,minimum width=1.3cm,minimum height=1.2cm, auto, very thick, },
        input/.style={ 
        draw,
        trapezium,
        minimum width=1.5cm,
        align=left,
        minimum height=1cm
    },
        }
        \node [block]  (start) {\textbf{Discover}\\ categorical \\combinations};
        \node[anchor=south,text width=2.4cm] (note1) at (-0.6,-1.3) {Sec. 2.1, 3.1, 3.2};
        \node [block, right =0.5cm of start, align=left, text width=2.6cm] (rec) {\textbf{Rank \& Recommend} \\ abnormal patterns};
        \node[anchor=south,text width=2.1cm] (note1) at (3.2,-1.3) {Sec. 3.3, 3.4};
        \node [block, right =0.5cm of rec, align= left, text width=2.2cm] (vis) {\textbf{Visualize \& \\ Interpret} rank ordering differences};
        \node[anchor=south,text width=2.1cm] (note1) at (6.3,-1.3) {Sec. 3.5};
        \path[draw,->] (start) edge (rec)
                       (rec) edge (vis);
\end{tikzpicture}
\vspace{-0.3in}
\caption{Computation Components of \emph{Helios}. Sec.2.3, Sec. 3.1-3.5 describe in detail \emph{Helios}'s holistic system design for SPM. Sec. 4 evaluates performance, discovery efficiency, and interpretability.}
\end{figure}
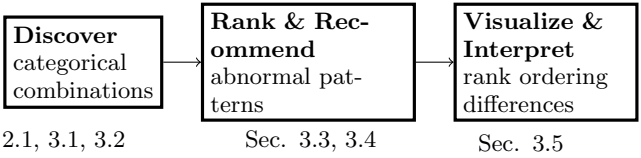

\begin{figure*}
\includegraphics[height=1.4in, width=7in]{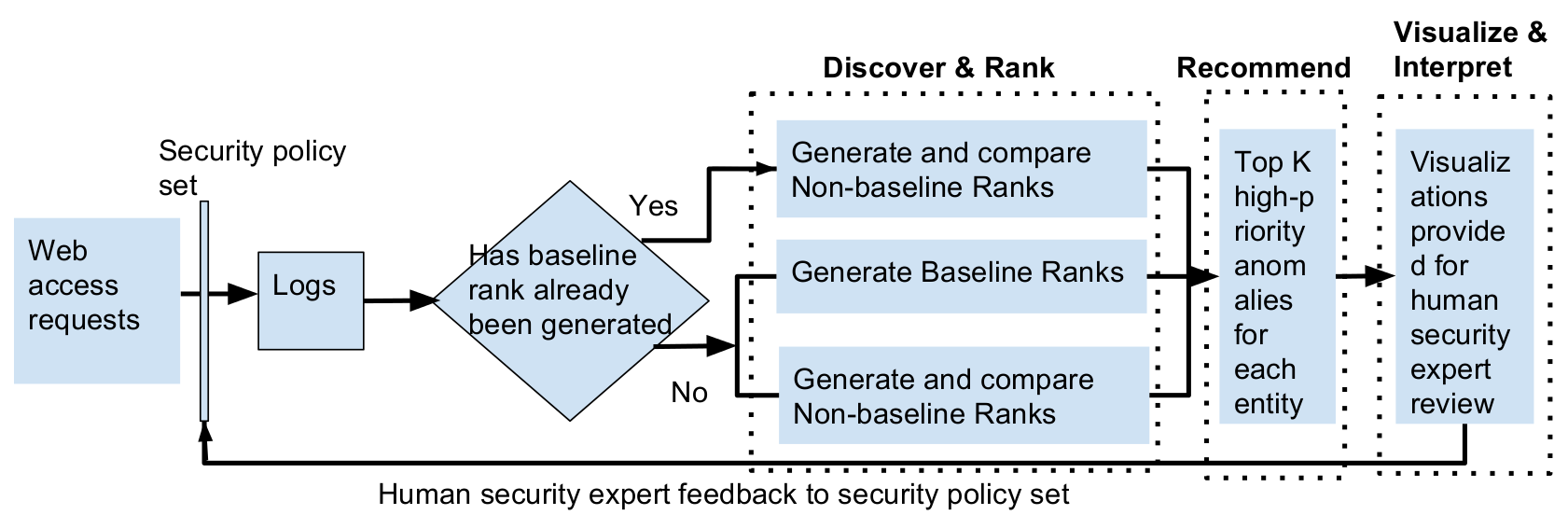}
\caption{SPM with \emph{Helios}: Discover, rank, recommend anomalies with human feedback to security policies}
\end{figure*}
\vspace{-0.02in}
The short E2E TTD makes \emph{Helios} very suitable for ongoing SPM: \emph{Helios} can quickly discover new categorical combinations which could be potential latent threats, and then provide interpretable visualizations(sec.4.2) to explain rationales in discovery process, helping human security experts make decisions. 

\emph{Helios}'s scalability potential is also suitable for large-scale SPM. Typically, in a moderately sized enterprise system, between $\approx 1.5 - 2.6 * 10^8$ entries are recorded in access logs each hour, and the total number of log entries grows roughly linearly in time. Moreover, the search space of categorical combinations is also quite large, while staying roughly constant as the time window widens. For instance, in log of a typical weekday, with only 6 common categories(Table 1), there can be roughly 
$200 \times 350 \times 50 \times 100 \times 100 \approx 3.5 \times 10^{10}$ unique categorical combinations, not counting $\approx 2.5 * 10^6$ IP addresses. 

\begin{table} 
\centering
\caption{Cardinalities of Categories}
\begin{tabular}{|c|c|} \hline
Category Name&Cardinality\\ \hline
Country & $\geq 200$ \\ \hline
Browser & $\geq 350$\\ \hline
Request Type& $\geq 50$\\ \hline
Customers& $\geq 100$\\ \hline
Content Type& $ \geq 100$ \\ \hline
IP Addresses& $ \geq 2.5 * 10^6 $ \\ \hline \end{tabular}
\end{table}

The rest of the paper is organized as follows. Section 2 clarifies regularly used terminologies and phrases we use in the paper, and overviews the background of previous efforts on building systems to automate certain parts of the SPM process. Section 3 describes the system design and discovery process in detail(Fig. 1 shows the design overview); section 4 lays out experiments setups for evaluating \emph{Helios} and results, and we conclude that \emph{Helios} indeed leads to more efficient SPM anomaly discovery. Section 5 concludes current work and introduces points of future research interest.  

\section{Overview}
We clarify terminologies, examine related work, and lay out challenges we in designing a system for discovering and recommending unknown anomalies.
\subsection{Terminologies}
To more clearly deliver our discussions in the following sections, we now qualitatively define certain terms we use, especially in the context of SPM and recommender systems. 
\begin{enumerate}[nosep]
\item[Def.1] \textbf{\emph{Mean reciprocal rank(MRR)}}: A longstanding metric in evaluating recommender and information retrieval systems\cite{Voorhees:mrr} and has several key advantages(Sec.3.2), MRR is mean value of the reciprocals of an entity's rank positions across different rank orderings. MRR of entity $c$ across $n$ rank orderings is defined as \vspace{0.1in}\\
\vspace{0.1in}
\centerline{
\begin{math} MRR_{c} \gets \frac{1}{|n|} \sum_{j=1}^{|n|}{\mathit{RR_{(c,j)}}} \end{math} 
} \\
where $RR_{(c,j)}$ is the reciprocal of $c$'s rank position in a rank ordering $j$ , where all existing entities in $j$ are sorted by a ranking function $F$ based on measurement metric $M$. In our experiments, we sort all rank orderings by decreasing order based on counts of access log entries. Fig.3 below shows steps to compute MRR \\
\vspace{-0.1in}
\begin{figure}[H]
\centering
\begin{tikzpicture}
 \tikzset{block/.style= {draw, rectangle, align=center,minimum width=1.5cm,minimum height=1cm},
        input/.style={ 
        draw,
        trapezium,
        minimum width=1.5cm,
        align=left,
        minimum height=1cm
    },
        }
        \node [block] (start) {Rank positions};
        \node [block, right =0.8cm of start] (rr) {Reciprocal ranks \\(RR)};
        \node [block, right =0.8cm of rr] (mrr) {MRR};
        \path[draw,->] (start) edge (rr)
                    (rr) edge (mrr);
\end{tikzpicture}
\caption{Compute MRR from a rank ordering} 
\end{figure}
\vspace{-0.2in}
An example of the computation process is as follows: in 5 different rank orderings, item $c$ has rank positions: 
$(2, 2, 10, None, 5)$. Then, RR for $c$ in these orderings are: $(\frac{1}{2}, \frac{1}{2}, \frac{1}{10}, None, \frac{1}{5})$. Finally, 
$MRR_{c} \gets \frac{1}{|5-1|}(\frac{1}{2} + \frac{1}{2} + \frac{1}{10} + \frac{1}{5}) = 0.325$. So in other words, MRR is the harmonic mean of rank positions. 
\item[Def.2] \textbf{\emph{Cold start}}: When the discovery process starts, no a priori information is given or inference can be concluded about users or items. The cold start problems we have seen in recommender systems research, are primarily starting without user information. \cite{}
\item[Def.3] \textbf{\emph{Categorical combination}}: A  length $n$ categorical combination of categorical labels, is an ordered tuple of length $n$, where each value in the tuple is a categorical value. Multiple access entries may map to identical categorical combinations. \\ 
For example, \textit{(Firefox, United States, text/html)}
is a length 3 categorical combination with ordered categorical labels from 3 categories in an access log: ``Browser'', ``Country'', and ``Content Type'', and many entries share this combination of browser, country, and content type in the access logs. 
\item[Def.4] \textbf{\emph{Security policy management(SPM)}}: Ongoing configurations of rules and policies for access and traffic control for servers, clients, services, and/or applications. A human security expert is involved to adjust and refine the configurations. Security services with SPM are known as ``managed security services'', e.g. managed WAF is a WAF with SPM.
\item[Def.5] \textbf{\emph{End-to-end time-to-discovery(E2E TTD)}}: A straightforward metric to measure efficiency of unknown anomaly discovery. The starting ``end'' is web access logs that record any traffic not blocked by WAF, the finishing ``end'' is the discovered anomalies. The time between starting to process web access logs and finishing searching for anomalies is the TTD. The shorter TTD is, the more efficient a system is in discovery new anomalies. 
\item[Def.6] \textbf{\emph{Supervision}}: In machine learning(ML) contexts, ``supervision'' qualitatively describes the amount of labeled data needed for training ML system. Needing only minimal amount of labeled data(Sec.3) to start the discovery process, \emph{Helios} is a minimally supervised system.
\end{enumerate}

\subsection{Related Work}
In recent years, there have been considerable efforts in high-profile security conferences on building systems to automate parts of the SPM process. Several notable works focus on Android SPM from different angles, and have built different systems with ML - especially natural language processing(NLP) - techniques in an effort to automate the SPM process. For example, \cite{peng2012using:risksofandroid} ranks risks of Android apps by na\"{\i}ve Bayes and a hierarchical Bayesian model, \cite{WHYPER:usenix2013:nlp:spm}identifies app permissions with a set of NLP techniques, \cite{usenix2015:rfforecase:incident} predicts network log data leak instances by a random forest, and \cite{usenix2015:semisup:EASEAndroid} constructed a ML system with a feedback loop for Android SPM based on a K-nearest neighbor process. Outside of the Android SPM sphere, \cite{makanju2009clustering} mined clusters out of event logs for system fault management purposes. On the other hand, recommender systems have succeeded in large-scale consumer-facing applications \cite{Gomez-Uribe:scalenetflix, Linden:amazonitembased, davidson2010youtube}, specializing in predicting and recommending items to users' preferences. Despite such success of recommender systems and a conspicuous call to build more efficient and automated SPM processes\cite{Landwehr:2008keynote}, using recommender systems for discoveries in efficient SPM has not been explored in publications. The venues we have searched range from dedicated one such as RecSys\cite{recsys:website}(for recommender systems) and top four security conferences\cite{secref:ms, secref:zhou, secref:gu}, and next to none was done to use recommender system for building more efficient, interpretable, hence more usable SPM system at large-scale. 

\subsection{Challenges}
Discovering previously unseen, therefore unknown anomalies for SPM is more challenging than classic anomaly detection and IDS designs. Classic anomaly detectors, WAF, and IDS raise alerts when certain pre-defined rules or certain priors are violated\cite{Steinwart:anomclass,Ali:anomclass:threshold,Eskin2002:decisionbound, wressnegger2013close, ding2012intrusion:measurement:ids}. However, such ``define and classify'' design is limited in capability to discover unknown anomalies, because they face four major challenges: \\
\\ 
\textbf{Ch.1}:\emph{Real-world logs after WAF and IDS filtering also contain latent potential threats}. Certain web access events can more easily evade detections than others do\cite{waf:xss:difficult, waf:sqlinjection, waf:sqlinjection:bypasswaf, xu2016automatically}, and it is often challenging trying to discover those potentially malicious access events from the wider context of presumed ``benign'' events in the WAF-filtered logs. Our purpose of building \emph{Helios} is precisely to address this ``finding the unknown unknowns'' problem in SPM. \\
\textbf{Ch.2}:\emph{Large-scale networks produce large number of access log events}, and the logs after filtering are still voluminous. Manual analyses of the access logs by human security experts to find new latent threats is time-consuming and impractical. Hence, an automated or semi-automated system is needed for post-filter logs. In Fig. 1, we illustrate computational components and non-computational system designs for a semi-automated human-in-the-loop SPM process with \emph{Helios} handling the large scale. Experiments in Sec.4 further shows scalability performance and discovery efficiency. \\
\textbf{Ch.3}:\emph{Scarce knowledge of rules and priors beforehand for latent potential threats}. Unlike anomaly detectors, WAF, and IDS with pre-defined normality, rules, and/or priors, it is difficult to extract information or infer knowledge about what the latent threats could be. In fact, even security experts often do not know what could potentially be found in access logs. In Sec. 3.1, 3.2, and 3.3, we go through details of discovering and recommending anomalies in a double-sided cold-start scenario, where neither user- nor item-related information is supplied in access log. \\
\textbf{Ch.4}:\emph{Assumption of stationarity}: SPM is an ongoing process because both benign behaviors and latent threats keep evolving with new software, applications, services, and vulnerabilities. Pre-defined rules and priors are likely lagging behind the evolution process\cite{huang2011adversarial}. In Sec. 3.1, we describe a method of discovering baselines from current access logs in a bootstrapping manner. \\

Meanwhile, despite successful large-scale applications \cite{Linden:amazonitembased, Gomez-Uribe:scalenetflix}, two major approaches to recommender systems - collaborative filtering \cite{Sarwar:item1, Deshpande:item2} and content-based\cite{Melville:contentbased,Balabanovic:contentbasedfab} systems - both normally assume availability of implicit data, user metadata, and similarity measurement in the input \cite{Sarwar:item1, Shi:ltrcf, WeiKarLeSmo08:cofirank,Hu:implicit, Basu:infoused, Balabanovic:contentbasedfab}, which web access logs do not have(\textit{cf.} Sec.1). For example, widely-used common log formats specified by Apache and W3C\cite{logformat:w3c, logformat:apache} only include fields such as client user agent or browser, IP address, request, and request status, and even for sophisticated log formats, information needed by traditional recommender systems is still not available. For example, in a moderately sized enterprise network we work with, its customized and fairly sophisticated log format has these categories: \\

\textit{Unique ID, Source Entity, HTTP/HTTPS, Device, Device Operating System, 
Content Type, User Agent, Country, Request Status, Client IP, Content Length, 
Request URL, Request Type, Host, Proxy, Date, Hour, UNIX Epoch} \\
\begin{figure}[H]
\centering
\includegraphics[height=1.4in, width=3.4in]{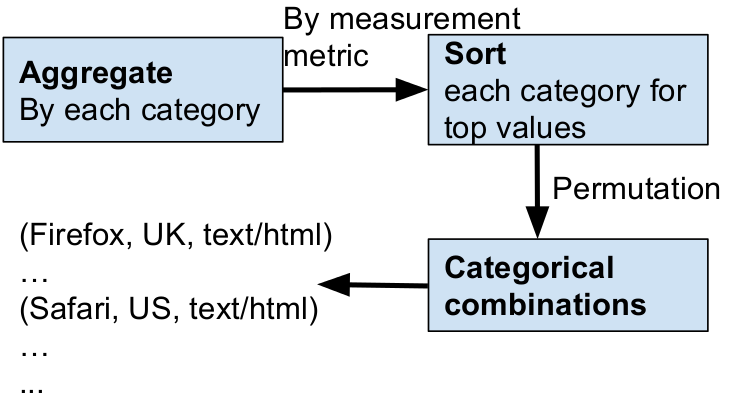}
\caption{Discover: Generate Baseline - Details} 
\end{figure}
Clearly, none of the categories in the log indicates human experts' preferences and prior knowledge, nor metadata and meaningful numerical variables for common recommender systems to construct ground truths, before even starting the recommendation process. 

In addition to the lack of user metadata and general prior knowledge, because new anomalous patterns are unknown and unseen, human security experts often do not even know what new unknown abnormalities to look for. This poses a double-sided cold start problem for recommender systems: there is no information about ``users''(security experts), and there is little information about the ``items''(abnormalities in access logs), either. Needing more information than what is available in access logs and from security experts, common recommender systems' approaches do not address the problem of unknown anomaly discovery.

\emph{Helios} fills this gap, by efficiently discovering and recommending unknown anomalies. Without the usual information available on users or items, \emph{Helios} bootstraps to discover, rank, recommend, and visualize salient anomalies for security experts. It also handles the heavy traffic load and number of categorical combinations to search through for discovery and ranking, which addresses a key pain point in today's SPM process(Sec.1). 

\section{RECOMMENDER SYSTEM FOR DISCOVERING \& RANKING ANOMALIES}
\subsection{Discover: Generate Baseline}
The double-sided cold start problem(Sec.2.3) is obvious in access log $L$: users do not have knowledge about unknown anomalies, and entries in $L$ gives no information about unknown anomalies either. Between the two difficult scenarios, extracting user-related knowledge is much less meaningful and practical: after all, an anomaly discovery and recommender system should find unknown anomalies for security expert users, not waiting for users to tell it what to look for. 
Hence, the only practical way to alleviate the double-sided cold start, is finding a way to extract item-related information from $L$. Given the vast number of entries in $L$(Sec. 1), and the formats of $L$ heavily use categorical values(Sec. 2.3), we propose a method to bootstrap baseline categorical combinations from $L$. After an example(\textit{cf.} Fig.4), we will give a more formal representation.

Suppose there are $m$ distinct categories $A_1, A_2,\ldots, A_m$ in $L$. Let $m=3$, where $A_1$ = ``Browser'', $A_2$ = ``Country'', and $A_3$ = ``Content Type''. $M$ is the number of entries logged in $L$, $F$ is sorting numerical values in descending order, and we take the top 2 values in each category. Therefore, $r_1$ is the rank ordering of all browsers descendingly by each browser's number of entries in $L$, and $A_{1,(3)} \gets r_1[:2]$ is the set of the 2 browsers with the largest numbers of entries logged in $L$. 

To find the top 2 values in each category, there are 3 steps(Fig.4). First, for each of the 2 categories, we aggregated all entries in the log to count the number of entries with every value within each category. Second, we sorted and ranked each category in descending order according to count of entries, hence we now have 3 rank orderings: $r_1$, $r_2$, and $r_3$, then we take 2 highest ranked values in each. For example, the values can be: \textit{Firefox, Safari}(Browser); \textit{US, UK}(Country); \textit{text/html, image/jpeg}(Content type). Finally, without loss of generality, suppose the length 3 categorical combinations will be in the form of ordered tuple \textit{(Browser, Country, Content Type)}. With 2 values for permutation in 3 categories, there can be $2^3=8$ categorical combinations: \textit{(Firefox, US, text/html), (Firefox,US,image/jpeg), (Firefox, UK, text/html), (Firefox,UK, image/jpeg),(Safari,US,\\text/html), (Safari,US,image/jpeg), (Safari,UK,text/html), (Safari,UK,image/jpeg)}. These 8 tuples form $S_{B}$, baseline categorical combinations discovered from $L$.
  
For a more formal representation, we let $n_j = \mathbf{card}(A_j), \forall 1 \leq j \leq m$ indicate cardinality of $A_j$, which is simply the number of distinct categorical values\footnote{We use $\mathbf{card}$ to indicate cardinality of a set} within the category $A_j$. 
As we already saw in example(Table 1, Sec.2.1), there could be $\prod \limits_{j=1}^m n_j$ unique categorical combinations, because of permutation $S_{all} = \{A_1 \times A_2 \ldots \times A_m\,\}$ that generates all categorical combinations, is Cartesian product of all $A_j$. Let $M$ be the measurement metric(s), $F$ be the ranking function, $p$ the number of values to take from each category\footnote{Here we set $p_{i} = p_{j}$, $\forall 1 \leq i,j \leq m$ and $i \neq j$, but there is absolutely no such restriction while generating baseline categorical combinations. To generalize, there are $\prod \limits_{j=1}^m p_j$ baseline categorical combinations.}, a more formal representation of our method to discover baseline categorical combinations is:

This process of discovering baseline categorical combinations address two challenges in Sec. 2.3. \textbf{Ch.3}(``cold start'') is a major challenge for recommender systems, and \textbf{Ch.4}(``assumption of stationarity'') is a well-known concern for ML-based security applications\cite{huang2011adversarial}. Without user- nor item-related information(e.g. preference, feedback or similarity), and without labeled data as input to train, \emph{Helios} discovers baselines from access log in a bootstrapping manner. Further, the baseline is drawn from \emph{current} access log, as opposed to using information from previous access logs, therefore this baseline can be considered near real-time, and will changes as access log changes. 

\subsection{Discover: Generate Non-baseline}
If a categorical combination does not belong to $S_{B}$ - the returned result from Sec.3.1 - it is a non-baseline categorical combination. Continuing the example in Sec.3.1, any categorical combination that is not one of the 8 tuples specified in the example, is considered non-baseline: If an ordered tuple of length 3 has the form \textit{(Browser, Country, Content Type)}, and the browser is not Firefox or Safari, or the country is not US or UK, or any content type tis not \textit{image/jpeg} or \textit{text/html}, the tuple is one of non-baseline categorical combinations. \textit{(Firefox, UK, text/plain)} is a non-baseline categorical combination, for instance. The 3 categories in $L$ could generate $\approx 350 \times 200 \times 100 \approx 7 \times 10^{6}$ unique categorical combinations, and other than the 8 tuples specified, all of the $7 \times 10^{6}$ categorical combinations are non-baseline. 

More formally, $S_{B}$ is the set of baseline categorical combinations discovered from log $L$, and $S_{all} = \{A_1 \times A_2 \ldots \times A_m\,\}$ is all unique categorical combinations in the form of length $m$ ordered tuples. $S_{nb}$, the set of non-baseline categorical combinations, is $S_{NB} \gets \{A_1 \times A_2 \ldots \times A_m\} \setminus S_{B}$, which is any categorical combination that is not in $S_{B}$.
\subsection{Rank: Compute Rank Statistics}
With baseline and non-baseline categorical combinations already discovered from $L$, we show in Fig.5 the next stages of ranking and recommending top $K$ most abnormal categorical combinations. We first compute MRR from baseline categorical combinations, and RR from non-baseline ones. Then we apply a distance function to measure the distances between MRRs and RRs for different categorical combinations, and rank such distances in descending order. Finally, those $K$ categorical combinations that come on top of the rank ordering are recommended as most abnormal categorical combinations to human experts for SPM. 

\begin{figure}[H]
\centering
\includegraphics[height=1.4in, width=3.4in]{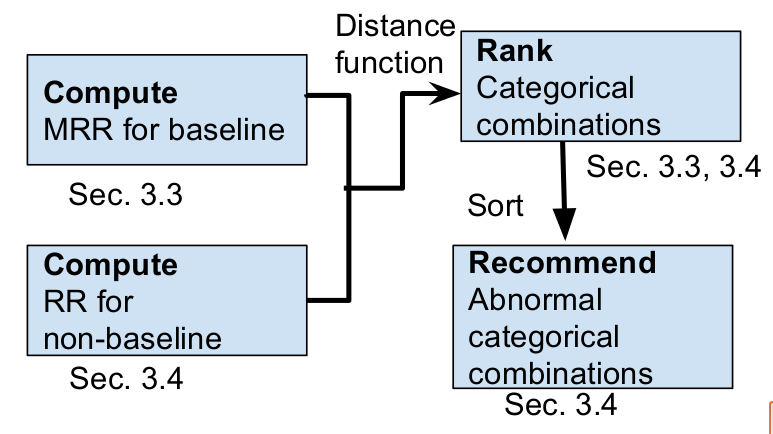}
\caption{Rank \& Recommend - Details} 
\end{figure}
\vspace{-0.18in}
As a discovery and recommender system, \emph{Helios} customizes its recommendations at the entity level. A typical example of ``entities'', is the group of end customers of network-based service providers such as content delivery networks and financial transaction processing gateways, which serve multiple customers' websites. As seen in Fig.2, access requests to customers' websites are filtered by security policies, and then recorded in access logs - where customers' websites are often identified as ``source entity'' in log formats. In the rank and recommendation stage, \emph{Helios} identifies different sets of most abnormal categorical combinations for different entities, and we will continue to describe the process with the same example setting in Sec.3.1-3.2. 

Take two customers codenamed ``SK'' and ``WN'' from 200 customers in $L$ as examples, and based on the same $M$(number of entries logged in $L$), $F$(sorting in descending order), we find that in each of $L$'s subsets which satisfies a baseline categorical combinations(e.g. all \textit{(browser, country, content type)} are set to \textit{(Firefox, UK, text/html)}), their rank positions among all customers, are $R_{SK, baseline}= (38, 22, 45, None, \\None, None, 37, 26)$, and $R_{WN, baseline} = (2, 1, 1, 3, 17, None,\\None, None)$, respectively. By Def.1, SK's and WN's MRRs calculated from the baseline categorical combinations, are listed in Table 2. By interpretation of MRR(Def. 1), SK's expected mean rank is $\approx$ 31, and WN's $\approx$ 2. 
Now consider two non-baseline categorical combinations, $s_{nb,1}$ and $s_{nb,2}$, from which SK's and WN's ranks and RRs are recorded in Table 2. Finally, let the distance function $D$ be L-1 distance, so we compute absolute values of differences between SK's MRR and RRs of SK's rank positions in all non-baseline categorical combinations. Results are shown in the last 2 rows of Table 2. Computing rank statistics continues on to all customers, thier MRRs, and their L-1 distances between MRRs and all non-baseline categorical combinations where these entities have rank position, and final results of L-1 distances are stored in a data structure indexed by entity and non-categorical combinations to prepare for the recommendation stage(Sec.3.4).
\vspace{-0.2in}
\begin{table}[H]
\centering
\caption{Rank: Compute Rank Statistics Examples}
\begin{tabular}{|l|l|l|} \hline
\textbf{Name}& \textbf{SK}& \textbf{WN} \\ \hline
$MRR_{*}$& 0.032& 0.578 \\ \hline
Expected rank& 1/0.032$\approx$31& 1/0.578$\approx$2\\ \hline
$R_{(*, nb,1)}$& 4& 10  \\ \hline
$RR_{(*, nb,1)}$& 1/4 = 0.25& 1/10 = 0.1\\ \hline
$R_{(*, nb,2)}$& 25& 50\\ \hline
$RR_{(*, nb,2)}$& 1/25 = 0.04& 1/50 = 0.02\\ \hline
$d_{(*, nb,1)}$& |0.032 - 0.25|=0.218& |0.578 - 0.1|=0.478\\ \hline
$d_{(*, nb,2)}$& |0.032 - 0.04|=0.008& |0.578 - 0.02|=0.576\\ \hline
\end{tabular}
\end{table}
There are 3 advantages of using RR and MRR $c_i$ instead of simple ranks or normalized ranks(\textit{cf.} Sec.2.1). 
\begin{enumerate}[nosep]
\item[\textbf{Adv.1}] Both RR and MRR are bounded: RR, MRR \begin{math} \in \left(0, 1\right]. \end{math} Same range for all entities being ranked, easy to measure and compare across rank orderings.
\item[\textbf{Adv.2}] MRR, as mean of multiple RRs, measures the magnitude of difference between one group of ranks and another group of ranks for the same entity. 
\item[\textbf{Adv.3}] By comparing one entity's MRR computed from the group of baseline ranks, and the same entity's RR in individual non-baseline ranks, we can measure how far each rank in an individual rank ordering(measured by RR) is from the expected mean rank(measured by MRR). Because each rank ordering is indexed by a categorical combination, the magnitude of differences is measurement for how far individual categorical combinations are from baseline ones
\end{enumerate}

\begin{algorithm}
\caption{Rank: Compute Rank Statistics}
    \SetAlgoVlined
    \SetInd{0em}{0.6em}
    \SetKwInOut{Input}{Input}
    \SetKwFor{ForEach}{forall}{do}{endfall}
    \setcounter{AlgoLine}{0}
    \setlength{\intextsep}{0pt}
    \Indm
    \Input{Access log $L$, baseline categorical combinations $S_{B}$, non-baseline categorical combinations $S_{NB}$}
    \Indp
    \nl initialize $D$, an empty indexable and sortable data structure \newline
    \tcp*[f]{for example, a Python list} \newline
    \nl\ForEach{$c_j$ $\in C$}
    {
      \tcp*[h]{Compute baseline MRR for entity $c_j$} \newline
      \nl $MRR_{j} \gets \frac{1}{\mathbf{card}(S_{B})}\sum_{q=1}^{\mathbf{card}(S_{B})}{\mathit{RR_{(q,j)}}}$ \newline
      \tcp*[f]{$RR_{(q)}$ is indexed by $s_{q}$} \newline 
     \nl \ForEach{$s_{nb} \in S_{NB}$}
      {         
         \tcp*[h]{append $D$ with the L-1 distance between $RR_{(nb)}$ and $MRR_{j}$} \newline
         \nl \begin{math} D \gets D \frown \left(\mathbf{d_1}(RR_{(nb)} - MRR_{j})\right) \end{math} 
       }
      }
      \nl\Return{$D$}
\end{algorithm}     
Algorithm 2 shows a more generalized formal representation: $\exists$ $C = \{c_1, c_2,\ldots, c_n\}$, a class of entities in $L$, and for some of $L$'s subsets $L_{(i)} = \{L \mid (A_1, A_2,\dots, A_m) = s_{i}\} \neq \emptyset$, where $s_{i}$ is a length $m$ categorical combination and $s_{i} \in S_{all}$, and $C_{(i)}$, a subset of $C$, is in $L_{(s,i)}$ . By measurement metric $M$ and ranking function $F$, within $L_{(i)}$, $C_{(i)}$ can form a rank order of its members: $R_{(i)} \gets F(C_{i}, by = M)$. Hence, it follows that RR of all $C_{(i)}$'s members is $RR_{(i)} \gets 1/R_{(i)}$, where $c_j$'s RR is $RR_{(i,j)}$. It then follows that if $s_{i} \in S_{B}$, for any $c_j \in C$, the MRR of $c_j$ is simply $MRR_{j} \gets \frac{1}{\mathbf{card}(S_{B})}\sum_{q=1}^{\mathbf{card}(S_{B})}{\mathit{RR_{(q,j)}}}$.

\subsection{Recommend: Find Most Abnormal Categorical Combinations}
When recommending most abnormal $K$ categorical combinations for each entity in $C$, we only make two simple assumptions:  
\begin{enumerate}
\item[\textbf{Asp.1}] For arbitrary $c_i$, MRR computed from baseline categorical combinations $S_{B}$ are considered normal for $c_i$, and $MRR_i$ would be largely preserved across categorical combinations considered normal for $c_i$. 
\item[\textbf{Asp.2}] For non-baseline categorical combinations $s_{nb,j}, s_{nb,q} \in S_{NB}$, $s_{nb,j} \neq s_{nb,q}$,  
an entity $c_i$, if L-1 distances: \\
\centerline{
\begin{math} \mathbf{d_1}(RR_{i,s_{nb,j}} - MRR_{i}) > \mathbf{d_1}(RR_{i,s_{nb,q}} - MRR_{i}) \end{math}, 
}
then for $c_i$, categorical combination $s_{nb,j}$ is more abnormal than categorical combination $s_{nb,q}$. \\
Similarly, if for another $s_{nb,t} \in S_{NB}$, if L-1 distances: \\
\centerline{
\begin{math} \mathbf{d_1}(RR_{i,s_{nb,t}} - RR_{i,s_{nb,j}}) < \mathbf{d_1}(RR_{i,s_{nb,t}} - RR_{i,s_{nb,q}}) \end{math}, 
} 
then for $c_i$, categorical combination $s_{nb,t}$ is more similar to categorical combination $s_{nb,j}$.
\end{enumerate} 
\setlength\itemsep{0em}

By \textbf{Adv.3} in Sec. 3.3 and the two assumptions, the recommender system regards that for an entity $c_i$, those non-baseline categorical combinations producing the largest L-1 distances between baseline categorical combinations, are regarded as most highly abnormal categorical combinations. For example, in Table 2, according to the last two rows recording L-1 distances between SK's and WN's MRRs to their RRs in two non-baseline categorical combinations, $s_{(nb,1)}$ is more abnormal for customer entity SK than $s_{(nb,2})$, while $s_{(nb,2)}$ is more abnormal for customer entity WN than $s_{(nb,1)}$. 
Suppose that a data structure $D_{SK}$ which stores and indexes SK's L-1 distances by categorical combinations, the following process finds top $K$ most abnormal non-baseline categorical combinations for SK. First: sort $D_{SK}$ values in descending order. Then, extract top $K$ non-baseline categorical combinations associated with top $K$ L-1 distances. This system workflow is described in Algorithm 3.

\begin{algorithm}
\caption{Rank \& Recommend Categorical Combinations}
    \SetAlgoVlined
    \SetInd{0em}{0.6em}
    \SetKwInOut{Input}{Input}
    \SetKwFor{ForEach}{forall}{do}{endfall}
    \setcounter{AlgoLine}{0}
    \setlength{\intextsep}{0pt}
    \Indm
    \Input{Integer $K$, data structure $D$ from Algorithm 2 with all L-1 distances indexed by entity $c_i$ and non-baseline categorical combinations $s_{nb}$}
    \Indp      
     \nl Initialize $DS$, an indexable data structure to store results \newline
      \tcp*[h]{For example, a Python dictionary or array} \newline
      \nl\ForEach{$c_i \in C$} 
      {
      \nl $D_{c_i} \gets sort(D_{c_i}, order = descending)$ \newline
      \tcp*[h]{Find top $K$ categorical combinations with largest L-1 distances to $MRR_{i}$} \newline
       \nl $s_{(nb,i)} \gets D_{c_i}[:K]$ \newline
       \tcp*[h]{Append $DS$ with top K categorical combinations, indexed by entity $c_i$} \newline
       \nl $ DS \gets DS \frown s_{(nb,i)}  $ 
     }
      \nl\Return{$DS$}
\end{algorithm}

\subsection{Visualize \& Interpret}
A natural follow-up from human security experts, after seeing the $K$ most abnormal categorical combinations, would be asking ``why'' \emph{Helios} finding these particular categorical combinations most abnormal. Indeed, recommendations without interpretable rationales as support evidence, are not very helpful in SPM: security experts need to understand, explain, and justify reasons, before making decisions on adjusting existing security policies. To address this need, \emph{Helios} provides visualizations based on RR, MRR, and rank orderings to help security experts understand why \emph{Helios} made such decisions. We give an example in Sec.4.3.

\section{EVALUATIONS}
The challenges we address with \emph{Helios} for SPM, are 1)to improve performance at large-scale applications, and 2)to boost discovery efficiency for unknown abnormal patterns in access logs(Sec. 2.3). To test \emph{Helios}'s performance and efficiency baseline, we ask the following research questions:
\begin{itemize}[nosep]
\item[\textbf{RQ1}] Does \emph{Helios} maintain \emph{reasonable E2E TTDs} at different scales?
\item[\textbf{RQ2}] Does \emph{Helios} improve efficiency for discovering abnormal categorical combinations in access logs, comparing to the largely manual process that many human security experts go through regularly? In other words, does \emph{Helios} \emph{shorten E2E TTDs} for human experts? If so, by how much?
\end{itemize}
In following evaluation experiments, we use large-scale real-world access logs with limited computing resources, to test for performance, potential for large-scale applications, and discovery efficiency. We chose ``customers'' to be the class of entities $C$, and 4 categories to generate categorical combinations: Category $A_1$ as ``Browser'', $A_2$ ``Country'', $A_3$ ``Content Type'', $A_4$ ``Request Type''. Hence, the categorical combination would be the length 4 ordered tuple of the form \textit{(Browser, Country, Content Type, Request Type)}. We also set measurement metric $M$ as number of log entries, ranking function $F$ as unweighed $M$ sorted in descending order, number of categorical values to take while discover baseline(Sec.3.1) as $p$ = 2, and top $K$=5 most abnormal categorical combinations for each entity. \\

\subsection{One: Performance Experiment} 
\begin{figure} 
\includegraphics[height=2.3in, width=3.4in]{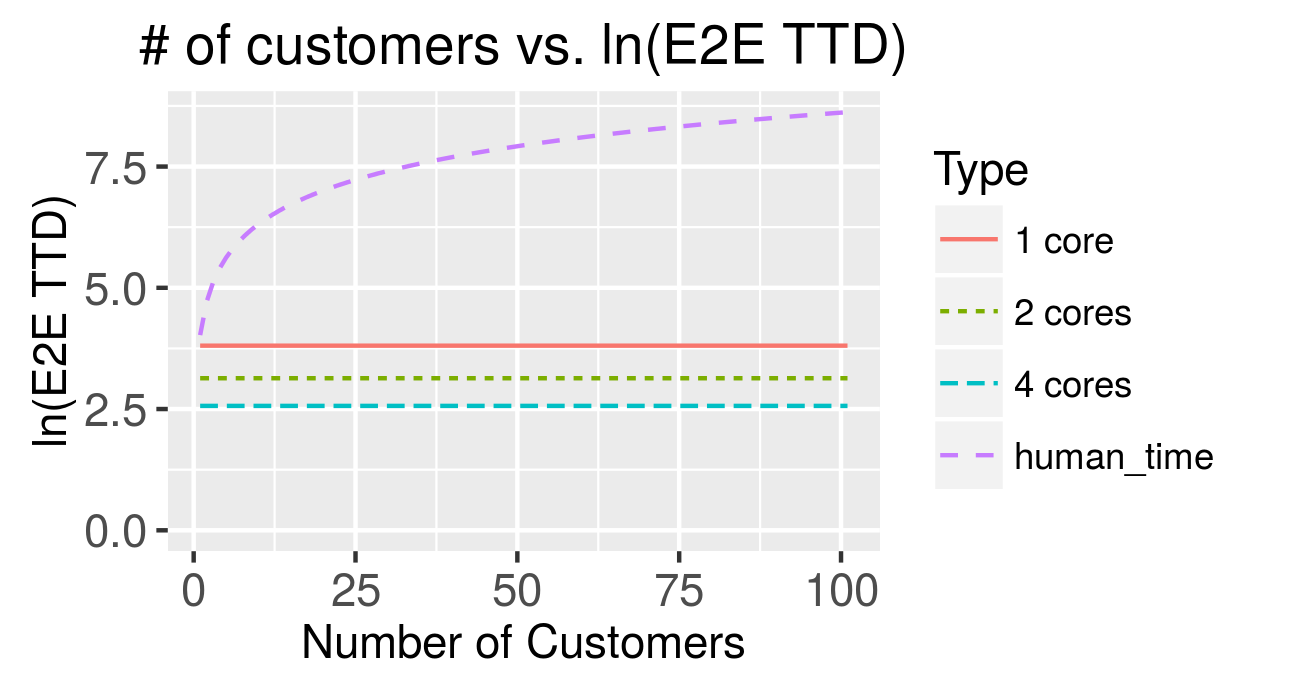}
\vspace{-0.1in}
\caption{ln(E2E TTD) for human and \emph{Helios}, as number of customers grows}
\vspace{-0.1in}
\end{figure}

We used a 45 GB 3-hour access log $L_{h1}$ of $6.5 * 10^8$ entries(retrieved without the ``hours'' mark), a 15 GB 1-hour log $L_{h2}$ of $1.01 * 10^8$ entries, and a 6 GB 1-hour log $L_{h3}$ of $8.3 * 10^7$ entries for testing performance benchmark. Any tools or platforms with reasonable distributed computing, storage, and support for indexable and sortable data structures are sufficient, and in the examples, we used Apache Spark on a local standalone cluster(a laptop with 4 GB working RAM and 4 2.30GHz processors). There are $\mathbf{card}(C)$ > 120 customers to be ranked in each log, and at least $\approx 200 \times 350 \times 50 \times 100 \approx 3.5 \times 10^{8}$ of categorical combinations to search through, as generated by the Cartesian product of $\{A_1, A_2, A_3, A_4\}$. We observed that, the cardinality of each category does not change dramatically according to the length of time window. This means that computing performance of Helios is not heavily impacted by bigger data volume from larger time windows. Specifics of the 3 sets are listed in Table 3.
\begin{table} 
\centering
\vspace{-0.1in}
\caption{Experiment Specifications}
\begin{tabular}{|l|l|l|l|} \hline
Category&${L_{h1}}$&${L_{h2}}$&${L_{h3}}$ \\ \hline
$\mathbf{card}(entries)$& $6.5 * 10^8$& $1.01 * 10^8$& $8.3 * 10^7$ \\ \hline
$\mathbf{card}(A_1)$& 400+& 350+& 350+ \\ \hline
$\mathbf{card}(A_2)$& 200+& 200+& 200+ \\ \hline
$\mathbf{card}(A_3)$& 160+& 150+& 120+ \\ \hline 
$\mathbf{card}(A_4)$& 200+& 50+& $\approx 50$ \\ \hline
{\textbf{E2E TTD}}& $\leq 60$mins& $\leq 55$mins& $\leq 50$mins \\ \hline
\end{tabular}
\end{table}
Note that although cardinality of log entries increases roughly linearly, the cardinalities of categories - especially $A_1$, $A_2$, $A_3$ - do not explode nearly as rapidly. We processed $L_{h1}$, $L_{h2}$, $L_{h3}$ each 3 times, and recorded E2E TTDs in Table 3. We checked for all categories' cardinalities for a typical 24-hour-window of estimated $\approx$400GB access log with 4.6 billion entries. The cardinalities are very similar to those of $L_{h1}$, meaning that likely, it can be processed within 60 minutes to produce top $K$ most abnormal categorical combinations to each entity. This means that prioritized recommendations can be generated for a week's worth of logs in less than an hour on a 4-processor laptop. \\

\subsection{Two: Discovery Efficiency Experiment} 
For this experiment, we used another one-hour access log of 13.2GB and $2 * 10^8$ entries in a cold start environment.
To evaluate the discovery efficiency, we ran the experiment on a standalone Apache Spark cluster: a desktop with 16 GB working RAM and 8 3.40GHz processors. We set the default parallelism to 4 and the number of executor to 1, and only change the number of executor cores as 1, 2, and 4. Under each core number, 3 experiments were run, and the E2E TTD are recorded in the table below. The TTDs are rounded up or down to the nearest minute, and we have found that given the same number of cores, the performance is remarkably consistent. We also asked a human expert - with assistance of his own tools - to start from the same dataset, and recorded the discovery time for comparison. Using his own tools and methods, our human expert started from the same access log as \emph{Helios} did, and attempted to find highly abnormal patterns. 
\begin{table}[H]
\centering
\caption{Experiment: Discovery Efficiency}
\begin{tabular}{|l|l|l|l|l|} \hline
\multicolumn{1}{|l|}{}& 1 core& 2 cores& 4 cores \\ \hline 
\multicolumn{1}{|l|}{\emph{Helios}}& 45min& 23min& 13min\\ \hline 
\multicolumn{3}{|c|}{Human Expert}& $\approx$ 110min(one entity) \\ \hline 
\end{tabular}
\end{table}

Our security expert spent $\approx$56 minutes processing access log data, and $\approx$54 minutes searching for potential abnormalities for one specific customer. The projection is that for each customer who demands SPM, he would spend $\approx$5 to 50 minutes to search for potential abnormalities. For \emph{Helios}, when the 3 computation components(Fig.1) are finished, each customer's top $K$ most abnormal categorical combinations are already stored in a data structure. Hence, more generally, as number of entities increase, E2E TTD for human experts to execute discovery and recommendation process for SPM scales roughly linearly. Meanwhile, \emph{Helios}'s E2E TTD remains largely constant. Fig.6 compares E2E TTD between human experts and \emph{Helios} as number of customer who needs SPM grows. 

\subsection{Three: Interpretability} 
Suppose that one of the most abnormal categorical combinations for the customer SK is \textit{(Mobile Safari,Singapore,\\text/plain,POST)}(Fig.8), and one of the baseline categorical combinations is \textit{(Chrome,United States,image/jpeg,GET)}(Fig.7). How does Helios decide that the condition in Fig.8 warrants additional human expert attention? With 2 typical figures, we describe how \emph{Helios} provides relevant visualizations to give straightforward interpretations to security experts on decision rationales(Sec.3.5).
\begin{figure} 
\includegraphics[height=2.2in, width=3.8in]{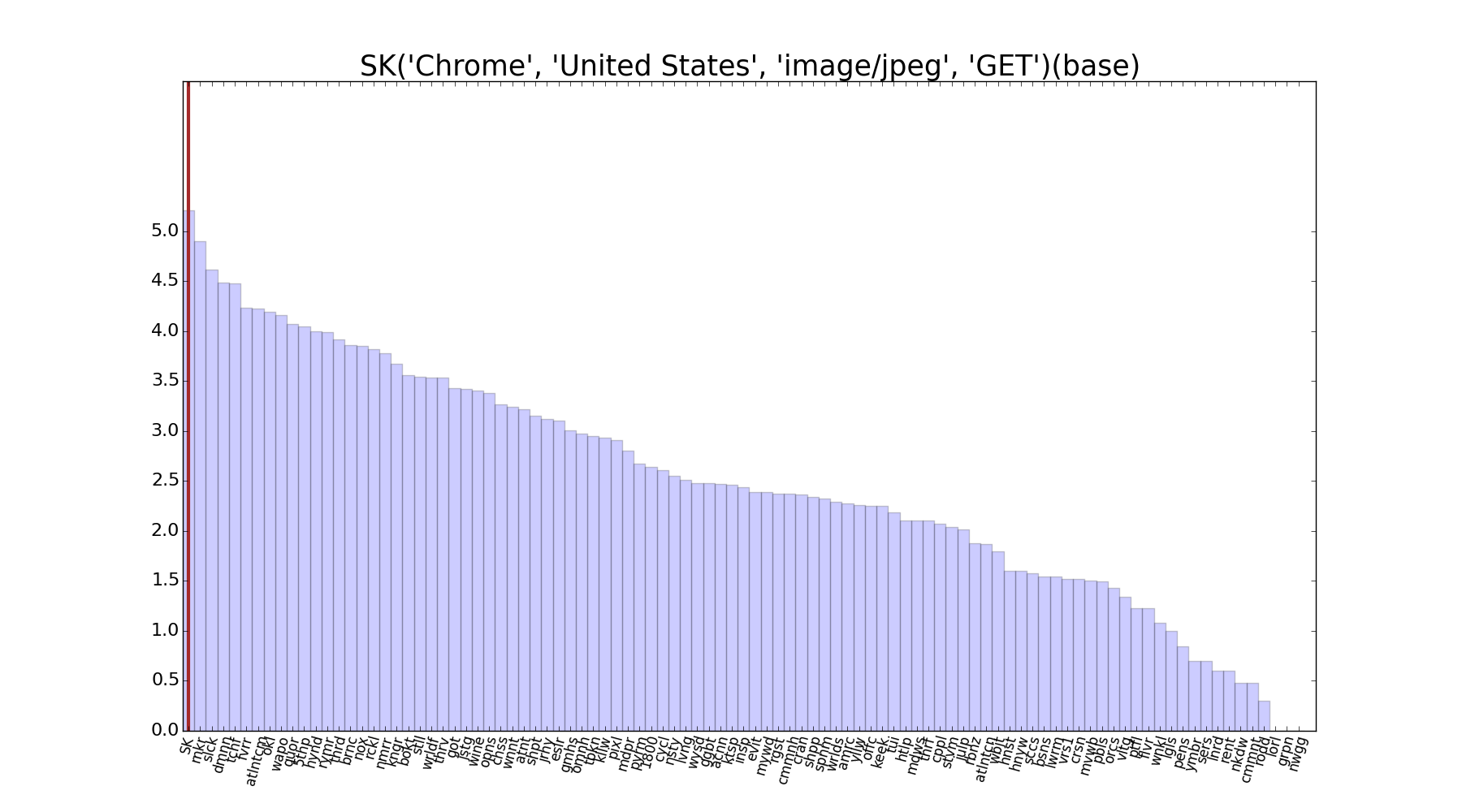}
\caption{SK's rank in one baseline categorical combination \textit{(Chrome,United States,image/jpeg,GET)}}
\end{figure}
\begin{figure} 
\includegraphics[height=2.2in, width=3.8in]{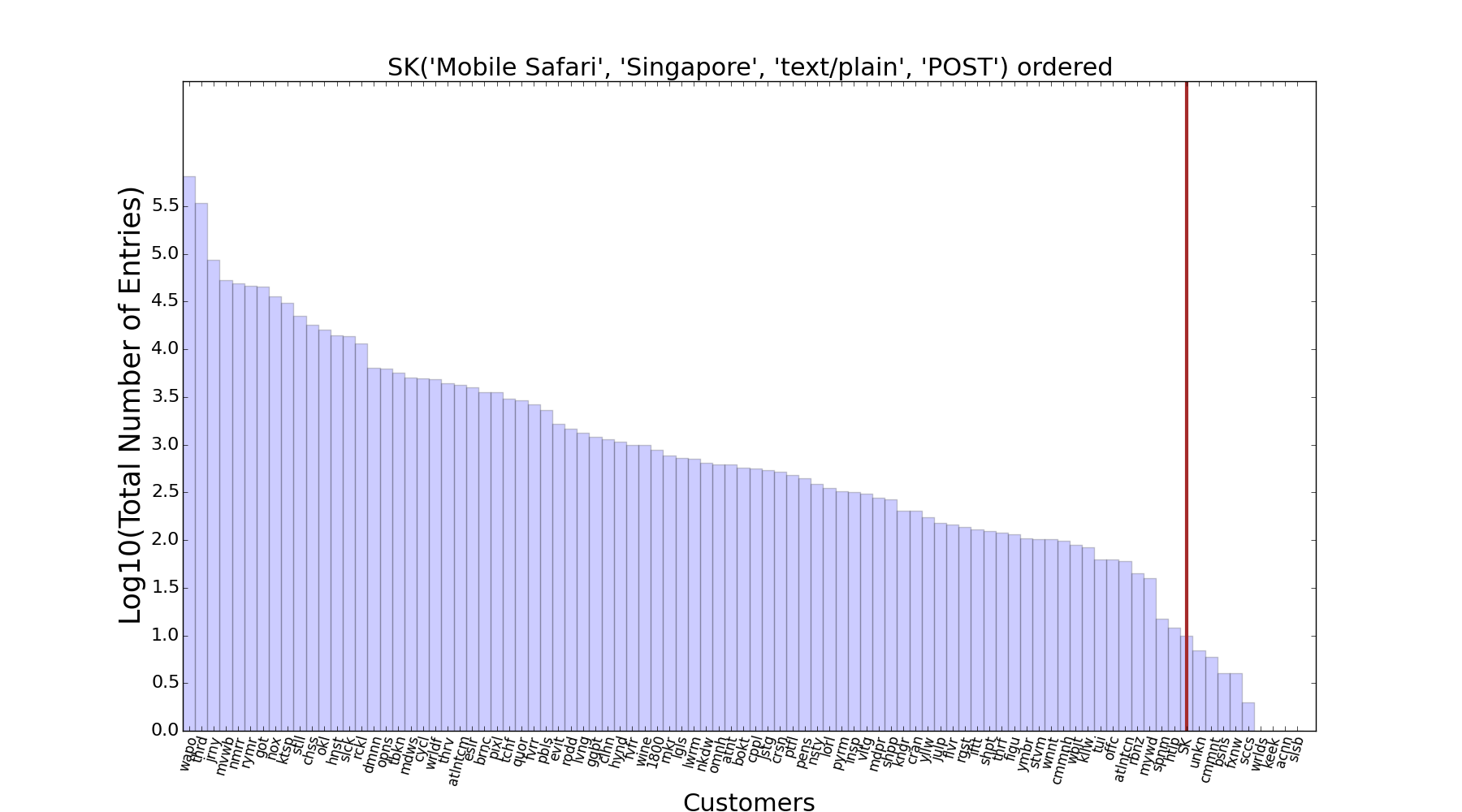}
\caption{SK's rank in one of the most abnormal categorical combination \textit{(Mobile Safari,Singapore,text/plain,POST)}}
\end{figure}
The thin vertical lines in Fig.7(on far left close to y-axis) and Fig.8 indicate customer SK's rank position among all other customers with access log events under each categorical combination. In Fig.7, the vertical line is close to the y-axis, indicating that under the specific baseline categorical combination, and based on the measurement metric $M$(count of log entries) and ranking function $F$(unweighed $M$ sorted in descending order), SK's rank is 1. In Fig.8, the vertical line is drastically shifted to the right, directly showing the visual difference that based on the same $M$ and $F$ but under different categorical combinations, SK's ranks drop dramatically and almost fall to the bottom of all customers. 

Fig.7 represents one of the baseline categorical combinations(We provide only one such baseline graph for illustration purpose. In practice, because there are multiple baseline categorical combinations, \emph{Helios} provides as many baseline graphs as the number of baseline categorical combinations. The only exception is that if under a certain combination, an entity has no events in the filtered access log, then the baseline graph is not provided.), while Fig.8 represents one of the most abnormal categorical combinations. 

\section{CONCLUSIONS AND FUTURE WORK}
We presented \emph{Helios}, a novel, non-standard recommender system for large-scale SPM: it efficiently discovers unseen hence unknown abnormal patterns from web access logs. Based on a generic minimally supervised method, \emph{Helios} does not rely on having user or item data beforehand that are commonly available to recommender systems. Using discrete categorical labels from access logs to build categorical combinations, \emph{Helios} offers a flexible and interpretable discovery engine for abnormal categorical combinations in access logs. Experiments showed that \emph{Helios} largely augments security experts' capabilities to discover unknown new latent threats at large scale, and our approach has four advantages:
\begin{enumerate}
\item \emph{Helios}' discovery and recommendation mechanism is intuitive and interpretable: constructing baseline categorical combinations is similar to building queries in a SQL-like grammar, and discrete categorical values are treated as content features.
\item \emph{Helios} is flexible: Users can customize baseline categorical combinations $S_B$, class of entities $C$, measurement metrics $M$, and ranking function $F$ based on availability of priors, and then use \emph{Helios} to discover and prioritize unknown and unseen abnormalities.
\item Minimally-supervised, \emph{Helios} learns to recommend anomalies from double-sided cold starts, where no information from users or items are available. \emph{Helios}' visualizations provide reasons of recommending certain categorical combinations as most anomalous.
\item \emph{Helios} scales well even with limited computing resources\\(Sec.4.1-4.2), and this makes it suitable for large-scale customized SPM applications.
\end{enumerate}

We plan to incorporate relevance feedback from the users to provide even more targeted results, as feedback can be treated as ``implicit feedback'' from the point of view of a content-based recommender system. Currently, our system deals with log data from specific lengths of time windows such as 1 hour or 1 week, and we plan to build a generative model for temporal pattern analysis, based on our efficient model as shown in Sec.4.2. Our further plan also includes metric learning, imputations of missing ranks, and grammar induction for semantic recommendations.


\end{document}